\documentclass[a4paper,11pt,openright]{article}
\usepackage{slashed}
\usepackage{cancel}
\makeatletter
\def\cleardoublepage{\clearpage\if@twoside \ifodd\c@page\else
 \hbox{}
 \vspace*{\fill}
 \begin{center}
 \end{center}
 \vspace{\fill}
 \thispagestyle{plain}
 \newpage
 \if@twocolumn\hbox{}\newpage\fi\fi\fi}
\makeatother

\def\a{{\zeta}}

\def\xx{{\xi}}
\def\ppsi{\Psi}

\def\wl{w_{\lambda}}

\def\formdif{{\alpha}}

\def\bC{{\bf C}}

\def\bZ{{\bf Z}}

\def\CC{{\cal C}}

\def\TT{{\cal T}}

\def\wh{\widehat}

\def\wl{w_{\lambda}}

\def\formdif{{\alpha}}

\def\bC{{\bf C}}

\def\bZ{{\bf Z}}

\def\CC{{\cal C}}

\def\TT{{\cal T}}

\def\wh{\widehat}

\def\half{{\scriptstyle{1 \over 2}}}
\def\interior#1{\setbox1=\hbox{$#1$}\rlap{$#1$}\kern0.4\wd1\raise1.1\ht1%
\hbox{$\scriptstyle \circ$}}

\def\boxit#1#2{\setbox1=\hbox{\kern#1{#2}\kern#1}%
\dimen1=\ht1 \advance \dimen1 by #1 \dimen2=\dp1 \advance \dimen2 by #1
\setbox1=\hbox{\vrule height\dimen1 depth\dimen2\box1\vrule}%
\setbox1=\vbox{\hrule\box1\hrule}%
\advance \dimen1 by .4pt \ht1=\dimen1 \advance \dimen2 by .4pt \dp1=\dimen2
\box1\relax}
\def\endprf{\raise .5ex\hbox{\boxit{2pt}{\ }}}

\def\ifundefined#1{\expandafter\ifx\csname#1\endcsname\relax}

\def\beq{\begin{equation}}
\def\endq{\end{equation}}
\def\eeq{\end{equation}}
\def\beqa{\begin{eqnarray}}
\def\bea{\begin{eqnarray}}
\def\endqa{\end{eqnarray}}
\def\eea{\end{eqnarray}}

\def\D{{D}}
\def\X{\tilde{X}}

\def\dateline{\today}

\date{\dateline}

\def\s{{t}}
\def\W{{W}}
\usepackage{amssymb}

\title{de Sitter symmetry of Neveu-Schwarz spinors}
\author{Henri Epstein$^{1}$ and Ugo Moschella$^{2}$\\
$^{1}$Institut des Hautes Etudes Scientifiques (IHES), \\ 35, Route de Chartres,  91440 Bures-sur-Yvette, France\\
$^{2}$Universit\`a dell'Insubria, DiSat, Via Valleggio 11, 21100 Como\\
and INFN sezione di Milano, Italia}
\date{\dateline}

\begin{document}
\maketitle
\begin{abstract}
We study the relations between Dirac fields living on the 2-dimensional 
Lorentzian cylinder and the ones living on the double-covering of the 2-dimensional de Sitter manifold, 
here identified as a certain coset space of the group $SL(2,R)$. 
We show that there is an extended notion 
of de Sitter covariance only for Dirac 
fields having the Neveu-Schwarz anti-periodicity and construct the relevant cocycle. 
Finally, we show that the de Sitter symmetry is naturally 
inherited by the Neveu-Schwarz massless Dirac field on the cylinder.
\end{abstract}

\section{Introduction}
There are two spin structures and,  
correspondingly, two possible choices  of  free  fermions on the  (Minkowskian or Euclidean) cylinder. 
According to the terminology of string theory, fields which are periodic 
are called Ramond (R) fields, fields which are anti-periodic are called Neveu-Schwarz (NS) fields.
The conformal mapping of the Euclidean cylinder onto the punctured complex plane changes this situation:  here Neveu-Schwarz fields are periodic 
and Ramond fields are anti-periodic, showing that the NS condition is the more natural one,  the one which fits better with conformal invariance (see e.g \cite{difrancesco, gsw, shenker}).

In this paper we examine another kind of conformal mapping: 
the map of a portion of the Minkowskian cylinder onto the two-dimensional de Sitter spacetime and the transformation properties of 
quantum Dirac fields under such mapping.  
The fields obtained in this way on the two-dimensional de Sitter universe have no a priori reason to fit with the de Sitter symmetry of the manifold.
We will show however that  the canonical quantum Dirac field  subject to the NS anti-periodicity  condition does enjoy a certain natural form of de Sitter covariance. Also, the inverse image of that representation shows that NS  canonical Dirac fields carry a hidden de Sitter symmetry already on the cylinder. On the other hand, the field which satisfies the Ramond periodicity condition transforms covariantly only under the (compact) subgroup of spatial rotations while the boosts  are broken.

The construction goes as follows: after recalling the canonical quantizations of either periodic (R) or anti-periodic (NS) Dirac fields on the Minkowskian cylinder,  we consider quantum Dirac fields on the two-dimensional de Sitter universe. 
There are two possible ways of writing a Dirac equation on that manifold. The first equation, recalled in Section \ref{FDI},  is the standard Dirac-Fock-Ivanenko equation constructed in terms of the spin connection (see e.g. \cite{birrel}). The second equation, described in Section \ref{DDir}, arises in a construction due to Dirac himself \cite{dirac} that is directly related to a particular unitary representation of the de Sitter group \cite{bargmann}.
In the four-dimensional case the relation between the two equations has been elucidated by F. G\"ursey and T. D. Lee \cite{gursey}.
The two-dimensional case that we are considering here is more subtle for topological reasons. 
It turns out that a transformation matrix field  {\em \`a la}  G\"ursey and  Lee interpolating between the two Dirac fields only exists 
on the double covering of the (two-dimensional) de Sitter manifold - identified in Section \ref{coset} as the coset space of the spin group $SL(2,R)$ over the abelian subgroup $A$. 
This construction allows to transfer the representation of the de Sitter group naturally  associated to the Dirac's equation of Section \ref{DDir} to the fields that are solution of the Dirac-Fock-Ivanenko equation. 
The de Sitter covariance is now expressed in terms of a cocycle that we explicitly construct in Section \ref{cocycle}. The result is that Neveu-Schwarz fields  on the double covering of the de Sitter hyperboloid are fully de Sitter covariant  while Ramond fields are only rotation invariant. 

In the last Section we exhibit the conformal transformation relating  de Sitter covariant field on the de Sitter manifold and the massless NS Dirac field on the cylinder, which therefore carries a hidden de Sitter symmetry.  On the other hand the same mapping applied to the Ramond field produces a field which only has the rotation invariance. It is well known that fields do not have to be single-valued in a non simply connected space \cite{isham,isham2}.
Our result helps to understand the naturalness of the NS boundary condition in the study of 2-dim QFT models \cite{difrancesco,zittartz,manton}.

The formalism and the results presented here may be useful in the study of various conformal field theories on the two-dimensional de Sitter universe. 
In an accompanying paper we will present a construction of the de Sitter -Thirring model.

\section{The Dirac equation on the 2-dimensional Minkowskian cylinder} \label{diracil}
Consider a Minkowskian cylinder parametrised with a time coordinate $x^0 = \s$ and a periodic spatial coordinate $x^1 = \theta $ (with $-\pi <\theta <\pi$);  
 the metric coincides with the Minkowski metric:
\bea
&& ds^2 = d\s^2- d\theta^2, \label{cylmetric}
\eea
apart from the periodicity of  the $\theta$ coordinate. The Clifford anti-commutation relations
\bea
&& \gamma^{a}\gamma^{b}+\gamma^{b}\gamma^{a}=2\eta^{ab}, \\ && \eta_{ab} =  {\mathrm {diag}}(1,-1), \ \ a,b=0,1,
\label{Cliffordcyl}
\eea
may be realised by the following choice of gamma matrices:
\beq
\gamma^{0}=\left( \matrix{0&1\cr 1&0\cr}\right) ,\quad
\gamma^{1}=\left( \matrix{0&1\cr -1&0\cr}\right) .
\label{rapprcyl}
\eeq
There are two inequivalent spin structures on the cylinder which correspond to two different monodromies for the spinor fields:
periodic spinors: $$(R)  \ \ \ \  \ \psi(\s, \theta +2\pi) = \psi(\s, \theta) ;$$  and
anti-periodic spinors: $$(NS) \ \ \ \ \  \psi(\s, \theta +2\pi) = -\psi(\s, \theta).   $$
Let us consider now  the massless Dirac equation on the cylinder:
\beq
i \gamma^a \partial _a \psi = 0. \label{dirac}
\eeq
There are at least two canonical quantum fields that solve this equation and that correspond to the  above two monodromies.
To fix notations and for the reader's convenience we recall  them here. 
They are constructed in terms of two sets of creation and annihilation operators
\beq
a_j(p),\ b_j(p),\ a_j^*(p),\ b_j^*(p),\ \ \ \ j = 1,\ 2,\ \ \ 
p\in\bZ,\ \ p\ge 0\ ,
\label{z.1}\endq
acting in a Fock space with vacuum $\Omega$ 
\beq
a_j(p) \Omega = 0,\ \ \ b_j(p) \Omega = 0,
\label{z.3}\endq
and obeying the canonical anti-commutation relations:
\beq
\{a_j(p),\ a_k^*(q)\} = \delta_{jk}\delta_{pq}\ ,\ \ \ 
\{b_j(p),\ b_k^*(q)\} = \delta_{jk}\delta_{pq}\ .
\label{z.2}\endq
Every other anticommutator vanishes.

\vskip20pt
\noindent {\bf Ramond canonical field.} In terms of  the light-cone variables 
$$u = t+\theta, \ \ \ \ \ v = t-\theta$$ 
the components of the quantum Ramond-Dirac  field are written as follows:
\begin{eqnarray}
&&\psi^{\makebox{\tiny R}}_1(x) = \psi^{\makebox{\tiny R}}_1(u) = {1 \over 2\sqrt{\pi}} (a_1^*(0) +b_1(0))
 +
{1\over \sqrt{2\pi}} \sum_{p > 0} 
(a_1^*(p)e^{ipu} + b_1(p)e^{-ipu}), \cr
&&\psi^{\makebox{\tiny R}}_2(x) = \psi^{\makebox{\tiny R}}_2(v) ={1 \over 2\sqrt{\pi}}(a_2^*(0) +b_2(0)) +
{1\over \sqrt{2\pi}} \sum_{p > 0} 
(a_2^*(p)e^{ipv} + b_2(p)e^{-ipv}) .
\label{z.100a}
\end{eqnarray}
(There is however a certain arbitrariness in the choice of the canonical operators relative to the zero modes. We agree  here to the standard choice) .

$\psi^{\makebox{\tiny R}}$ is a univalued function on the cylinder.
The nonvaninshing 2-pt functions are given by
\begin{eqnarray}
 w_1(x, y) = (\Omega,\ \psi^{\makebox{\tiny R}}_1(x){\psi^{\makebox{\tiny R}}_1}^*(y)\Omega) 
= {1\over 2\pi}\Big [ \half + \sum_{p>0} e^{-ip(u-u')} \Big ]= -{i \over 4\pi }\cot\left(\frac{u-u'-i0}2\right)   \\   \cr
\label{z.101}
 w_2(x, y) = (\Omega,\ \psi^{\makebox{\tiny R}}_2(x) {\psi^{\makebox{\tiny R}}_2}^*(y)\Omega) = {1\over 2\pi}\Big [ \half + \sum_{p>0} e^{-ip(v-v')} \Big ]= -{i \over 4\pi }\cot\left(\frac{v-v'-i0}2\right) 
\label{z.102ter}\end{eqnarray}
In the usual matrix form (with $\overline {\psi} = {\psi}^+\gamma^0$)
\begin{eqnarray}
 (\Omega,\  \psi ^{\makebox{\tiny R}}(x){\overline {\psi}^{\makebox{\tiny R}}}(y) \Omega)  
= -\frac{i}{4\pi} \left(
\begin{array}{cc}
 0 & \cot\left(\frac{1}{2} (u-u' )\right) \\
   \cot \left(\frac{1}{2} (v-v' )\right) & 0 \\
\end{array}
\right).\label{z.102a}
 \end{eqnarray} 
\vskip20pt
\noindent {\bf Neveu-Schwarz canonical field.} Neveu-Schwarz Dirac fields are obtained by summing over half-integer momenta; one consequence of the shift $p\to p+1/2$   is that  there are no more zero modes and that all the modes  enter in the field expansion with the same normalization:
\begin{eqnarray}
\psi^{\makebox{\tiny NS}}_1(u) &=
{1\over \sqrt{2\pi}} \sum_{p \ge 0} 
(a_1^*(p)e^{ipu+iu/2} + b_1(p)e^{-ipu-iu/2}),\cr
\psi^{\makebox{\tiny NS}}_2(v) &=
{1\over \sqrt{2\pi}} \sum_{p \ge 0} 
(a_2^*(p)e^{ipv+iv/2} + b_2(p)e^{-ipv-iv/2}).
\label{z.100}
\end{eqnarray}
Here $\psi^{\makebox{\tiny NS}}$  is a bivalued function on the cylinder.
The 2-pt functions is now given by
\begin{eqnarray}
 (\Omega,\   \psi ^{\makebox{\tiny NS}}(x){\overline {\psi}^{\makebox{\tiny NS}}}(y) \Omega)  
= - \frac { i }{4\pi }\left(
\begin{array}{cc}
 0 &\frac{1}{\sin \left(\frac1{2}({u-u'}) \right)}\\
\frac{1}{\sin \left(\frac1{2}({v-v'}) \right)}& 0 \\
\end{array}
\right).\label{z.102}
 \end{eqnarray} 
\subsection{Scalar ancestors}
In correspondence with the  above construction one may also introduce two scalar massless fields. 
The field that one would naturally get by the method of canonical quantisation, defined on the cylinder itself, would be plagued by an infrared divergence which arises from a diverging normalisation of the zero mode.
This divergence may be cared by replacing the (infinite) constant zero mode by a normalised zero mode growing with time 
Performing this regularisation one gets a canonical quantum field but the price to pay is that positive-definiteness has been lost; the two point function is 
\begin{eqnarray}
 (\Omega,\  \varphi ^{\makebox{\tiny R}}(x){\varphi^{\makebox{\tiny R}}}(y) \Omega) = \wh \D_0(x-y) = 
-{ix^0-iy^0\over 4\pi} +{1\over 4\pi} \sum_{p\in \bZ \setminus \{0\}}
{e^{-i|p|(x^0-y^0)+ip(x^1-y^1)} \over |p|} \cr  = 
-\frac1{4 \pi } \log \left(-\frac{1}{4} \sin \left(\frac{1}{2} (u-u')\right)
\sin \left(\frac{1}{2} (v-v')\right)\right)
 \label{aqq}
 \end{eqnarray}
so that 
\begin{eqnarray}
 (\Omega,\  \psi ^{\makebox{\tiny R}}(x){\overline {\psi}^{\makebox{\tiny R}}}(y) \Omega)  =
 i \gamma^a \partial _a \wh \D_0(x-y) .
 \label{scalar1}
 \end{eqnarray} 
The two-point function of the 
second scalar field is constructed by shifting the series in Eq. (\ref{aqq}) to half integer momenta and suppressing the zero modes as follows:
\begin{eqnarray}
 (\Omega,\  \varphi ^{\makebox{\tiny NS}}(x){\varphi^{\makebox{\tiny NS}}}(y) \Omega) = \D_{\makebox{\tiny NS}}(x-y) = 
{1\over 4\pi} \sum_{p\in \bZ }
{e^{-i|p+1/2|(x^0-y^0)+i(p+1/2)(x^1-y^1)} \over |p+1/2|} \cr  = 
\frac1{4 \pi } \log \left(-\cot \left(\frac{1}{4} (u-u')\right)
\cot \left(\frac{1}{4} (v-v')\right)\right)
 \label{aqqns}
  \end{eqnarray}
 so that 
\begin{eqnarray}
 (\Omega,\  \psi ^{\makebox{\tiny NS}}(x){\overline {\psi}^{\makebox{\tiny NS}}}(y) \Omega)  =
 i \gamma^a \partial _a \D_{\makebox{\tiny NS}}(x-y) .
 \label{scalar1bis}
 \end{eqnarray} 

\section{The Dirac equation on the two-dimensional de Sitter spacetime}\label{FDI}
Consider now a (classical or quantum) massless Dirac spinor having the following form: 
\begin{equation}
\psi =( \cos \s) ^{-\frac 12}\phi. 
\end{equation}
The spinor $\phi$ itself solves a closely (conformally) related Dirac equation
\beq
i\gamma^a \partial _a \phi + \frac {i } 2 \tan \s \, \gamma^0   \phi =  0 \label{diracconf}
\eeq
which may be interpreted as the massless  Dirac (Fock-Ivanenko) equation on the de Sitter manifold. Let us  consider indeed the two dimensional de Sitter spacetime represented as the one-sheeted hyperboloid 
\beq
dS_{2}=\left\{ X \in {M}_{3}:\ (X^{0})^{2}-(X^1)^{2}-(X^2)^{2}=- 1 \right\} \label{ds}
\eeq
embedded in the three-dimensional Minkowski spacetime ${M}_{3}$ whose metric tensor is
\beq
\eta_{\alpha\beta} =  {\rm diag}(1,-1,-1) \label{3dimeta},\ \ \alpha,\beta = 0,1,2. \label{ambient}
\eeq
The following choice of coordinates 
\bea
X(t,\theta) = \left\{\begin{array}l
X^0=  r \tan \s \\
X^1 =  r   \sin \theta /\cos \s\\
X^2 =  r \cos \theta /\cos\s
\end{array}\right. \label{coorconf}
\eea
covers the complement of the lightcone in the ambient space: all the events which are  spacelike w.r.t to the origin $(0,0,1)$. In the above coordinates  
the de Sitter manifold ($r=R=1$) is manifestly conformal to a portion (i.e. $-\pi/2 <t<\pi/2$) of the Minkowskian cylinder:
\bea
&& ds^2 = \frac 1{\cos^2 \s} (d\s^2- d\theta^2);
\eea 
the interval of the ambient spacetime (restricted to $r=1$) is given by
\bea
&& (X-Y)^{2}= \frac{2\cos (\theta -\theta' )-2\cos (\s-\s')}{ \cos \s \cos \s' } .
\eea
The  components of the natural zweibein $e_a^i$ are 
\bea e^\s_0 = \cos \s, \ \ \ e^\s_1 = 0, \ \ \ e^\theta_0 = 0,\ \ \ \  e^\theta_1=\cos \s, 
\eea
and the curved-space matrices 
$ \alpha^i = e^i_a \gamma^a $ are simply  proportional to the flat space gamma matrices:
\begin{equation}
\alpha^\s = \left(
\begin{array}{cc}
 0 & \cos \s \\
 \cos \s & 0 \\
\end{array}
\right)= (\cos \s) \,\gamma^0, \ \ \ \ \ \alpha^\theta = \left(
\begin{array}{cc}
 0 & \cos \s \\
 -\cos \s & 0 \\
\end{array}
\right) = (\cos \s) \,\gamma^1. \label{alpha}
\end{equation}
In two-dimensions there is only one nonvanishing  component of the spin connection 
$
\omega_{\mu\, \!a b} = e_a^\nu \, \nabla_\mu e_{b\nu}.
$
In our example it is given by
$
\omega_{\theta 01} = -\omega_{\theta 10} = \tan \s.
$
Correspondingly 
\begin{equation}
\Gamma_\s=0, \ \ \ \Gamma_{\theta} =  \frac 14 [\gamma^0,\gamma^1] \omega_{\theta 01} = \tan \s \left(
\begin{array}{cc}
 -\frac{1}{2} & 0 \\
 0 & \frac{1}{2} \\
\end{array}
\right).
\end{equation}
Putting everything together, the massless Dirac equation on the de Sitter manifold is finally written as follows:
 \begin{eqnarray}
 i \alpha^i (\partial_i + \Gamma_i) \phi  = i \alpha^i \partial_i \phi +  i \alpha^\theta \Gamma_\theta \phi   = 
     i \cos t (\gamma^a  \partial_a  \phi +  \frac{i} 2 \tan \s\, \gamma^0  \phi)  =   0.
   \label{diraccurved}
 \end{eqnarray}
Eq. (\ref{diraccurved}) indeed coincides with Eq. (\ref{diracconf}). 
In the general case a mass term has to be added 
 \begin{eqnarray}   
 i \alpha^i (\partial_i + \Gamma_i) \phi - m\phi =0.  \label{diraccurved2}
 \end{eqnarray}
 and this of course breaks the conformal invariance.
 \section{Another equation by Dirac}
In this section we elaborate on another first order equation for a spinor on the de Sitter manifold which is due to Dirac himself \cite{dirac}.
Let us consider  the Clifford  anti-commutation relations
\beq
\gamma^{\alpha}\gamma^{\beta}+\gamma^{\beta}\gamma^{\alpha}=2\eta^{\alpha\beta}
\label{Clifford2}
\eeq
relative to the three-dimensional Minkowski  metric (\ref{3dimeta}). 
The following choice 
\beq
\gamma^{0}=\left( \matrix{0&1\cr 1&0\cr}\right) ,\quad
\gamma^{1}=\left( \matrix{0&1\cr -1&0\cr}\right) ,\quad
\gamma^{2}=\left( \matrix{i&0\cr 0&-i\cr}\right) 
\label{rappr2}
\eeq
provides a realisation of the above relations. The generators $L_{\alpha\beta}$  of the Lorentz group $SO_0(1,2)$ are given by 
\beq
L_{\alpha\beta}=M_{\alpha\beta}+S_{\alpha\beta},
\eeq
where
$ M_{\alpha\beta}=-i(X_{\alpha}\partial_{\beta}-X_{\beta}\partial_{\alpha})\label{genorb} $
and
$ S_{\alpha\beta}=-{i\over4}\lbrack\gamma_{\alpha},\gamma_{\beta}\rbrack\label{genspin} $
are respectively the 'orbital' and  the 'spinorial' parts of  $L_{\alpha\beta}$. 
In the coordinates (\ref{coorconf}) the orbital components 
\begin{eqnarray}
&& M_{01} = -i r \cos t \sin \theta \,  \partial_t - i  r\sin t \cos \theta\,  \partial_\theta \cr
&& M_{02} = -i  r \cos t \cos \theta \,  \partial_t+i  r \sin t \sin \theta\,  \partial_\theta \cr
&&M_{12} = -i r \partial_\theta \nonumber
\end{eqnarray}
obviously do not include derivatives w.r.t. the radial variable $r$. The differential operators $L_{\alpha \beta}$ are therefore tangential to the de Sitter hyperboloid (\ref{ds})
and have a well defined  action on functions and spinors defined just on it.
\label{DDir}

In the case under study ($s= 1/2$) the first Casimir operator takes the form\footnote{Eq. (\ref{casimir}) follows from the  identities
\bea
&& {{1}\over{2}}M^{\alpha\beta}M_{\alpha\beta}=-{\left({{1}\over{2}}\gamma_{\alpha}\gamma_{\beta}M^{\alpha\beta}\right)}^{2}-{{i}\over{2}}\gamma_{\alpha}\gamma_{\beta}M^{\alpha\beta}\ ,\label{MM2}\\
&& M^{\alpha\beta}S_{\alpha\beta}= -{{i}\over{2}}\gamma_{\alpha}\gamma_{\beta}M^{\alpha\beta}\ ,\label{MS2} \ \ \ \ \ 
{1\over 2}S^{\alpha\beta}S_{\alpha\beta}= {3\over 4}
\eea}
\beq
Q_{1} = - {1\over 2} L^{\alpha\beta}L_{\alpha\beta} ={ \left({1\over 2}\gamma_{\alpha}\gamma_{\beta}M^{\alpha\beta}\right)}^{2}+
i\gamma_{\alpha}\gamma_{\beta}M^{\alpha\beta}-{{3}\over{4}}. \label{casimir}
\eeq
and its eigenvalues ${1\over 4}+\nu^{2} $ are parametrized  by   a nonzero real number $\nu$   as described in Bargmann's classic paper \cite{bargmann}.
Since 
\beq
Q_{1} -\frac 14 ={ \left({1\over 2}\gamma_{\alpha}\gamma_{\beta}M^{\alpha\beta}+i\right)}^{2} \label{casimir2}
\eeq
following Dirac \cite{dirac} we may introduce another first order spinorial equation:
\beq
\left( iD+i+{ \nu} \right)\ppsi=  0\label{dSW}
\eeq
where we have set 
\beq
 iD = {1\over 2}    \gamma_{\alpha}\gamma_{\beta}M^{\alpha\beta} .
\eeq
\subsection{Spinorial plane waves}
The starting observation for solving Eq. (\ref{dSW}) is that 
the two-dimensional de Sitter - d'Alembert's operator can be  factorized as follows: 
$$
\left(
 \begin{array}{cc}
 \square & 0 \\
 0 &  \square  \\
\end{array}
\right) =  \cos ^2 t \left(
 \begin{array}{cc}
 \partial^2_t - \partial^2_\theta & 0 \\
 0 &  \partial^2_t - \partial^2_\theta  \\
\end{array}
\right) = -i D(i D+i). $$
This relation implies that 
\beq
\left( i{ D}+i+{ \nu} \right)\left( -i{D}+{ \nu} \right)\Psi=
(\square +{ \nu}^{2}+i{ \nu})\Psi.
\eeq
Given a two-component solution  of the scalar de Sitter Klein-Gordon equation 
\beq
(\square +{ \nu}^{2}+i{ \nu})\Psi = 0 \label{neiro}
\eeq
we may construct a solution of Eq. (\ref{dSW}) simply by applying the operator 
$\left( -i{D}+{\nu} \right)$ to it \cite{bartesaghi}. 
Let us recall how to find a suitable basis of scalar solutions for the  de Sitter Klein-Gordon equation \cite{bgm,bm}.
Given a nonzero light-like vector $\xi\in {M}_3$ and a complex number $\lambda \in \bC$ we construct the  homogeneous function
\begin{equation}
Z\mapsto (Z \cdot \xi)^\lambda.
\label{pwholo}\end{equation}
where $Z$  belongs to the  complex ambient Minkowski spacetime; this function is
holomorphic  in the tubes $T_\pm = {M}_3 \pm i V^+$, where $V^+$ is the future cone of the origin, and satisfies the massless Klein-Gordon
equation there. When restricted to the de Sitter universe it is holomorphic in the
tuboids  $\TT_\pm$ obtained by interesting the ambient spacetime tubes $T_\pm $ with the complex de Sitter manifold $dS_2^c$, 
and  satisfies the massive (complex) Klein-Gordon equation:
\begin{equation}
\left(\Box -\lambda(\lambda+1)\right) (Z \cdot \xi)^\lambda=0. \label{cdskg}
\end{equation}
The parameter $\lambda$ is here unrestricted, i.e. we consider
{ a complex squared masses} $m_\lambda^2=  -\lambda(\lambda+1)$.
The boundary values
\beq
(X\cdot \xi)_\pm^\lambda = \lim_{Z\in {\cal T}_\pm, \, Z \rightarrow X}
(Z\cdot\xi)^\lambda\ ,
\label{p.1}
\endq
are homogeneous distributions of degree $\lambda$ on $M_{3}$,
and their restrictions to $dS_2$, denoted with the same symbols,
are solutions of the de Sitter Klein-Gordon equation
\beq
\left(\Box +m_\lambda^2\right)(X\cdot \xi)_\pm^\lambda =0, \ \ \ \ \ \ \ \  \left(\Box +m_\lambda^2\right)(  X\cdot \xi)_\pm^{-1-\lambda} =0.
\label{p.2}\endq
All these objects depend in a $\CC^\infty$ way on $\xi$ and are entire in $\lambda$.

Coming back to the de Sitter-Dirac equation (\ref{dSW}), 
spinorial plane waves can therefore be written in terms of scalar plane waves  as follows \cite{bartesaghi}:
\beq
\left( -i{D}+{\nu} \right)(X\cdot \xi)_\pm ^\lambda v(\xi)\,  \label{solu}
\eeq
where $v= v(\xi) $ is a two-component spinor 
and the complex number $\lambda$ may take either of the following two values: 
\beq
{\lambda}_{1}=-i{ \nu},\ \ \ \ \ \ {\lambda}_{2}=-1+i{ \nu}.
\eeq
A straightforward calculation gives  that 
\bea
 -i{D}(X\cdot \xi)_\pm^\lambda  v(\xi) 
= {i\lambda}  (X\cdot \xi)_\pm^{\lambda-1} \slashed X w (\xi) - {i \lambda }    (X\cdot \xi)_\pm^{\lambda}v(\xi)
\eea
where as usual  $\slashed X= \gamma^\alpha X_\alpha$ and we set  $w(\xi) =  \slashed \xi v(\xi)$. The spinor $w(\xi)$ satisfies the condition ${\slashed \xi}w(\xi)=0$. In general, let us consider the linear equation
\beq
{\slashed \xi}u(\xi)=0.
\label{eqom2}
\eeq
For any given lightlike vector $\xi$ the unique solution  (apart from normalisation) of the above equation is 
\beq
u(\xi)= \frac 1 {\sqrt{2(\xi^0-\xi^1)}}\left( \matrix{ {\xi^0}-{\xi^1}\cr i{\xi_2}\cr }\right) =  
 \frac 1 {\sqrt{2}}\left( \matrix{ \sqrt{{\xi^0}-{\xi^1}} \cr i\sqrt{{\xi^0}+{\xi^1}} \cr }\right) \label{sspinn}
\eeq
Interestingly, this expression coincides 
with Cartan's original definition of a spinor, as described in his classic book (\cite{cartan} - Chapter 3). Defining the adjoint spinor in the  usual way
\begin{equation} 
{\overline u}(\xi)=u^{\dag}(\xi)\gamma^{0}
\end{equation}
there follows the completeness relation
\beq
 u(\xi)\otimes {\overline u}(\xi)=\frac 12\, {\slashed \xi }. \label{compl2}
\eeq
Let us suppose that $v(\xi)$ itself solves Eq. (\ref{eqom2}) and therefore that it has the form given in Eq. (\ref{sspinn}). We obtain the following (complete, unnormalised) set of spinorial plane wave solutions, labeled by the wave vector $\xi$ and the mass parameter $\nu$:
\begin{eqnarray}
&&
(X\cdot \xi)_\pm^{-1+i{ \nu}}u(\xi). \label{solI2} 
\end{eqnarray}
where $u$ is the spinor given in Eq. (\ref{sspinn}). It is now possible to build the two-point function of the quantum Dirac field $\ppsi_\nu$  
by superposing spinorial plane waves having the correct analyticity properties; these analyticity properties 
are there to replace of a true global spectral condition which is absent
in de Sitter Quantum Field Theory  \cite{bgm,bm,bem,bartesaghi}:
\begin{equation}
(\Omega,\  \ppsi_\nu (X_1)\overline \ppsi_\nu(X_2)\Omega) = \W_{\nu} (X_{1},X_{2}) 
= {c_{\nu}\over 2} \int_{\gamma}(X_{1}\cdot \xi)_-^{-1+i{ \nu}}(X_{2}\cdot \xi)_+^{-1-i{ \nu}}  \slashed \xi d\mu_{\gamma}(\xi)
\label{tpf2}
\end{equation}
and the canonical anticommutation relations fix the value of the constant:
\beq
c_{\nu}=\frac{\Gamma(1-i{\nu})\Gamma(1+i{ \nu})}{(2\pi)^{2}e^{\pi{\nu}}}.
\label{const2} \ \ \ \ \
\eeq
Since 
\bea
 \slashed X (-iD+\nu) (X\cdot\xi)^{-i\nu} = -\nu (X\cdot\xi)^{-1-i\nu}  \slashed \xi 
\eea
we get (in the complex analyticity domain ${\cal T}_-\times {\cal T}_+$)
\begin{eqnarray}
\W_{\nu} (Z_{1},Z_{2}) = -\frac {c_{\nu} }{2\nu}{\slashed  Z}_{2}(-i{D}_{2}+{\nu})
\int_{\gamma}(Z_{1}\cdot \xi)_-^{-1+i{ \nu}}(Z_{2}\cdot \xi)_+^{-i{ \nu}} \, d\mu_{\gamma}(\xi)
\label{tpf2massless} 
\end{eqnarray}
The integral at the RHS is a hypergeometric function of the invariant scalar product $\zeta = {Z_1}\cdot {Z_2}$ (see appendix \ref{BB}) 
and all the spinorial content is carried by the operator acting on it:
\begin{eqnarray}
\W_{\nu} (Z_{1},Z_{2})
= \frac {{\Gamma(1-i{\nu})\Gamma(i{ \nu})}}{{4\pi }} {\slashed  Z}_{2}(-i{D}_{2}+{\nu}) F\left(i\nu,\ 1-i\nu ;\ 1;\ {1-\zeta\over 2}\right )
\end{eqnarray}
Taking the massless limit $\nu\to 0$ we get 
\begin{eqnarray}
  \W_{0}(Z_{1},Z_{2})= - \frac {i} {4\pi} {\slashed  Z}_{2}
 + \frac {1} {4\pi} {\slashed  Z}_{2}(-iD_{2}) \sum_{n=1}^{\infty}
\frac{1} {n} \left({1-\zeta\over 2}\right)^n = \cr
= -\frac {i} {4\pi} {\slashed  Z}_{2}  
 + \frac {1} {4\pi} {\slashed  Z}_{2}(-iD_{2}) 
\log\left({1+\zeta\over 2}\right) 
=- \frac{i}{2\pi}
\frac{({\slashed Z}_{1}-{\slashed Z}_{2})}{(Z_{1}-Z_{2})^{2}}.\label{tpf20}
\end{eqnarray}
As a consequence of the above chain of identities we obtain a nice  
integral representation of the  two-point  function of  the massless field $ \ppsi_0$:
\begin{eqnarray}
 (\Omega, \ppsi_0 (Z_1)\overline \ppsi_0(Z_2)\Omega) =  
 \frac{1}{2\pi i}
\frac{({\slashed Z}_{1}-{\slashed Z}_{2})}{(Z_{1}-Z_{2})^{2}}= {1\over 8\pi^2}
\int_{\gamma}(Z_{1}\cdot \xi)^{-1}(Z_{2}\cdot \xi)^{-1}  \slashed \xi d\mu_{\gamma}(\xi)
\label{tpf2masslessbis}
\end{eqnarray}
valid for $Z_1\in {\cal T^-}$ and $Z_2\in {\cal T^+}$.

\section{Spin group and  de Sitter covariance}
Let us discuss now the de Sitter covariance of the Dirac field.
In our context the spin group (i.e. the double covering of $SO_0(1,2)$) is most usefully realized  as the matrix group
\beq
Sp(1,2)=
\bigl\{g\in SL(2,C) :\ \ \gamma^{0} g^{\dag}\gamma^{0}= g^{-1}\bigr \} .
\label{Sp}
\eeq
An element of $Sp(1,2)$ may be parametrised in terms of four real numbers $a,b,c,d$ subject to the condition  $a d+bc=1$ as follows:
\beq
g= 
\left(
\begin{array}{cc}
 a & i \,b \\
 i \,c & d \\
\end{array}
\right).
\label{sp12}
\eeq
As a subgroup of $SL(2,C)$, the group $Sp(1,2)$ is conjugated to $SL(2,R)$: the following element of  $SL(2,C)$
\begin{equation}
h=\left(
\begin{array}{cc}
 e^{\frac{i \pi }{4}} & 0 \\
 0 & e^{-\frac{i \pi }{4}} \\
\end{array}
\right)
\end{equation} 
does the job
\begin{equation}
h g h^{-1} = \left(
\begin{array}{cc}
 a & -b \\
 c & d \\
\end{array}
\right).
\end{equation}
The group $Sp(1,2)$ acts on the de Sitter manifold  by similarity transformations: 
\beq
\slashed X' =  g\slashed X g^{-1}.
\label{Sp2}
\eeq
The covering projection $g \rightarrow \Lambda(g)$ of $Sp(1,2)$ onto $SO_0(1,2)$ coherent 
with the above action is expressed as follows:
\beq
g\rightarrow {\Lambda(g)^\alpha}_\beta = {1\over 2}{\rm tr}(\gamma^{\alpha}g\gamma_{\beta}g^{-1}).
\label{lorentzmap}
\eeq
$\Lambda(g)$ is the (real) Lorentz transformation that directly relates $X'$ and $X$ 
 \begin{equation}
g\slashed X g^{-1} =  \cancel{\Lambda(g) X}. 
\end{equation}
The transformation law for spinors under the action of $Sp(1,2)$ is the standard one:
\begin{eqnarray}
\Psi^{\prime}(X)=g\Psi\bigr(\Lambda^{-1}(g)X\bigr)\label{trasI}, \ \  \ \ \ {\overline \Psi}^{\prime}(X)={\overline \Psi}\bigr(\Lambda^{-1}(g)z\bigr)g^{-1}.
\end{eqnarray}
The covariance of the two-point function (\ref{tpf2}) under the transformation (\ref{trasI}) can now be easily shown. In the massless case this is obvious
\beq
g\W_{0}(Z_{1},Z_{2})g^{-1}=\frac{1}{2\pi i}
\frac{g \,{\slashed Z}_{1}g^{-1}-g\,{\slashed Z}_{2}g^{-1}}{(Z_{1}-Z_{2})^{2}} 
= \frac{1}{2\pi i} \frac{\,{\cancel{\Lambda(g) Z}_{1}}-\cancel{\Lambda(g) Z}_{2}}{(Z_{1}-Z_{2})^{2}}.\label{tpf2ter0}
\eeq
In the general case, for any   
$g\in Sp(1,2)$ there holds the following chain of equalities:
\begin{eqnarray}
&&   g\W_{\nu}\bigl(\Lambda^{-1}(g)X_{1},\Lambda^{-1}(g)X_{2}\bigr)g^{-1}=  \cr \cr && =
{1\over 2} c_{\nu}\int_{\Gamma}\bigl(\Lambda^{-1}(g)X_{1}\cdot \xi\bigr)^{-1+i\nu}\bigl(\Lambda^{-1}(g)X_{2}\cdot \xi \bigr)^{-1-i\nu}
g{\slashed \xi}g^{-1}d\mu(\xi)
\cr & & =
{1\over 2} c_{\nu}\int_{\Gamma}\bigl(X_{1}\cdot \Lambda(g)\xi\bigr)^{-1+i\nu}\bigl(X_{2}\cdot \Lambda(g)\xi \bigr)^{-1-i\nu}
{\gamma^\alpha (\Lambda(g)\xi})_\alpha d\mu(\xi) \cr \cr
&&=  \W_{\nu}\bigl(X_{1},X_{2}\bigr)
\end{eqnarray}
The last step is a consequence of the Stokes theorem exactly as in the scalar case \cite{bm}.
\section{The symmetric space $Sp(1,2)/A $ as the double covering of the two-dimensional de Sitter spacetime.}
\label{coset}
In the following we need to  dwell a little longer on the group $Sp(1,2)$ and examine one of its coset spaces. The starting point is the Iwasawa decomposition $KNA$  of $Sp(1,2)$:
\begin{equation}
 g=k(\a)\,n(\lambda)\,a(\chi) = \left(
\begin{array}{cc}
\cos\frac \a 2 & i \sin\frac \a 2 \\
i \sin\frac \a 2  & \cos\frac \a 2 \\
\end{array}
\right) \left(
\begin{array}{cc}
 1 & i \lambda\\
 0 & 1 \\
\end{array}
\right) \left(
\begin{array}{cc}
e^{\frac \chi 2}& 0 \\
 0 &e^{-\frac \chi 2}\\
\end{array}
\right);\label{iwadec}
\end{equation}
the parameters $\a,\lambda$ and $\chi$ are related to $a,b,c$ and $d$ by the  following relations:
\begin{equation}
\cos\frac \a 2 =  \frac{a}{\sqrt{a^2+c^2}}, \ \ \  \sin\frac \a 2 =  \frac{c}{\sqrt{a^2+c^2}}, \ \ \ \lambda = a b-c d, \ \ \ e^{\frac \chi 2} = {\sqrt{a^2+c^2}}
\end{equation}
where $0\leq\a<4 \pi$ and $\lambda$ and $\chi$ are real. Note that $a$ and $c$ cannot be both 0 since $ad +bc = 1$.
The above Iwasawa decomposition provides a natural parametrization of the coset space $Sp(1,2)/A $ 
\begin{equation}
\tilde X (\lambda, \a)=k(\a)\,n(\lambda)= \left(
\begin{array}{cc}
 \cos \frac{\a}{2} & i \lambda  \cos  \frac{\a}{2} +i \sin  \frac{\a}{2} \\
 i \sin  \frac{\a}{2}  & \cos  \frac{\a}{2} -\lambda  \sin  \frac{\a}{2}  \\
\end{array}
\right)
\end{equation}
which is seen to be topologically a cylinder. 
The coset space $Sp(1,2)/A $ is a symmetric space. 
The group $Sp(1,2)$ acts on the coset space by left multiplication:
\begin{eqnarray}
g \, \tilde X (\lambda, \a)\rightarrow \tilde X (\lambda', \a') 
 \end{eqnarray}
It is useful to   describe the action of the subgroups separately. The case of a rotation $k(\alpha) \in K$
is of course the easiest one, it amounts simply to a shift of the angle $\a$:
\begin{eqnarray}
 \begin{array}{l} \lambda' (\alpha) =  \lambda ,\ \ \ \ 
 \a'(\alpha) = \a+\alpha.
\end{array}
\label{trasf1}
 \end{eqnarray}
The two other subgroups  give rise to slightly more involved transformation rules of the parameters $(\lambda,\zeta)$; an element $a(\kappa)$ of the abelian subgroup  $A$ gives
\begin{eqnarray}
\left\{ \begin{array}{l}  \lambda'(\kappa) = \lambda  \cosh \kappa+\sinh \kappa (\lambda  \cos \a+\sin \a), \\ 
\cot\frac {\a'(\kappa)} 2= e^{\kappa}\cot \frac \a 2 .
\end{array}\right.\label{trasf2}
 \end{eqnarray}
An element $n(\mu)\in N$  gives 
\begin{eqnarray}
\left\{ \begin{array}{l}  \lambda' (\mu)= {\lambda} \left(1 +\frac 12 {\mu ^2}\right)- \mu  \left( \lambda +\frac{\mu}2 \right) \sin \a
+{\mu } \left(1- \frac12 \lambda  \mu \right) \cos \a, \\ 
\cot \frac{\a'(\mu)} 2 =  \cot \frac \a 2 - \mu .
\end{array}\right.\label{trasf3}
 \end{eqnarray}The Maureer-Cartan form $dg \, g^{-1}$ gives to  the symmetric space $Sp(1,2)/A $ 
 a natural Lorentzian metric that may be constructed as follows (see e.g. \cite{mazur}). There exists a  inner automorphism of $Sp(1,2)$ 
\begin{equation}
g\rightarrow \mu(g) = - \gamma^2 g \gamma^2
\end{equation}
that leaves invariant the elements of the subgroup $A$. It may be used to construct  a map from the coset space $Sp(1,2)/A$ into the group $Sp(1,2)$:
\begin{equation}
g(\X) = g\mu(g)^{-1}=  -\X \gamma^2 \X^{-1} \gamma^2.
\end{equation}
This map in turn allows to introduce a left invariant Lorentzian metric on the coset space as follows: 
\begin{equation}
ds^2 = \frac 12 \makebox{Tr} (dg \, g^{-1})^2= -2 d\lambda d\a  -\left(\lambda ^2+1\right)d\a^2 \label{dscov}
\end{equation}
One may verify directly that:
\begin{enumerate}
\item The metric (\ref{dscov}) is invariant under the transformations (\ref{trasf1}), (\ref{trasf2}) and (\ref{trasf3}).
\item  The curvature is constant ($R=-2$) and the Ricci tensor is proportional to the metric:
\begin{equation}
R_{\mu\nu} - \frac 12 R g_{\mu\nu} = R_{\mu\nu} + g_{\mu\nu} =0
\end{equation}
\item The map $p: Sp(1,2)/A \rightarrow dS_2$
\begin{equation}
p : \tilde X (\lambda, \a) \rightarrow X(\lambda,\zeta)= \left\{
\begin{array}{l}
X^0= -\lambda  \\
X^1= \lambda  \cos \a+\sin \a \\
X^2= \cos \a-\lambda  \sin \a \\
\end{array}
\right.\label{iwa}
\end{equation}
is a covering map. 
\item Restricting the ambient spacetime metric (\ref{ambient}) to the de Sitter manifold with the coordinates (\ref{iwa}) gives again 
\begin{equation}
ds^2 = \left.\left({dX^0}^2-{dX^1}^2-{dX^2}^2\right)\right|_{dS_2} = -2 d\lambda d\a  -\left(\lambda ^2+1\right)d\a^2
\end{equation}
\end{enumerate}
{\em In conclusion: the symmetric space $Sp(1,2)/A = \widetilde{dS_2}$  may be identified with the double covering of the two-dimensional de Sitter universe.
The spin group  $Sp(1,2)$ acts directly on the covering space  $\widetilde{dS_2}$ as a group of spacetime transformations:}
\begin{equation}
 \X\to g\X
 \end{equation}
We were not able to find  the above identification in the (enormous) literature on the group $SL(2,R)$.

\section{Field dressing {\em \`a la} G\"ursey and Lee }
Since there are two apparently distinct Dirac's equations on the de Sitter manifold, namely Equation (\ref{diraccurved}) and Equation (\ref{dSW}), it is natural to ask whether there is a relation between them and what  it is.
In four dimensions, this question has been raised first by G\"ursey and Lee \cite{gursey} and they provided a way to build a 
bridge between the two  equations. The two-dimensional case is more  tricky (also more interesting)  because of its topological peculiarities. 
Following \cite{gursey} let us introduce three matrices
\beq
\beta^{\alpha}\equiv \left( \frac{\partial Y^{\alpha}}{\partial X^{\beta}} \right) \gamma^{\beta},
\eeq
where $Y^0 = t$, $Y^1 = \theta$ and $Y^2 = r$ are the coordinates (\ref{coorconf}).
In these coordinates the Minkowski metric of the ambient spacetime $M_3$ is written 
\begin{equation}
ds^2  = \frac {r^2}{\cos^2 \s} (d\s^2- d\theta^2) -dr^2 = r^2 g_{ij}dY^i dY^j - dr^2.
\end{equation}
Restricting to $r=1$  we find again the de Sitter metric and the following relations: 
\begin{eqnarray}
&& \{ \beta^{i},\beta^{j}\} =  \{\alpha^{i},\alpha^{j}\}=g^{ij }, \label{hhhh} \ \ \ i,j=0,1,\\
&&  \beta^{2} = -{\slashed  X}, \ \ \  \{ \beta^{i},\beta^{2}\} = 0.
\end{eqnarray}
Eq. (\ref{hhhh}) implies the existence of a matrix $S$ such that
\beq
\alpha^{i}=S\beta^{i}{S^{-1}}. \label{condd}
\eeq
A particularly convenient choice - see the discussion below - is the following particular element $S(\s,\theta)$ of the spin group $Sp(1,2)$:
\bea
S(\s,\theta)=\frac 1 {\sqrt{\cos \s}} \left(
\begin{array}{cc}
 {\cos \frac{t-\theta }{2}} & {i \sin \frac{t-\theta  }{2}}\\
 -{i \sin \frac{t+\theta  }{2}} & { \cos \frac{t+\theta  }{2}} \\
\end{array}
\right),
\eea
but any other matrix of the form 
$
S_f = f(t,\theta)S,
$
where $f(t,\theta)$ is an arbitrary function on the double-covering of the de Sitter spacetime $\widetilde{dS_2}$ is also permissible. 

Given a solution $\Psi$  of the Dirac Eq. (\ref{dSW}) the dressed spinor
\beq
\phi (t,\theta)= \frac 1 {\sqrt 2}  f(\s,\theta) S(t,\theta) (1- \slashed  X )\Psi (t,\theta)  \label{mapphi0} 
\eeq
solves the equation
\bea
  i  \alpha^t \left(\partial_t +\Gamma_\s \right)\phi   +    i  \alpha^\theta \left(\partial_\theta     +\Gamma_\theta \right)\phi   -   i  \alpha^i (\partial_i \ln f)  \phi - \nu \phi  = 0
 \label{eGurleetransf2n} 
\eea
(see Appendix \ref{GL}). 
Clearly, the arbitrary function $f(\s,\theta)$ can be reabsorbed by a gauge transformation. 
We may therefore set\footnote{If the function $f$ is also anti-periodic, i.e. if $ f(t,\theta+2\pi)=  -f(t,\theta)$ the matrix $S_f$ is well defined on the de Sitter hyperboloid ${dS_2}$ itself. However in this case we cannot solve the Dirac-Fock-Ivanenko equation with zero potential.} $f=1$ in Eq. (\ref{mapphi0}) and observe the coincidence of Eq. (\ref{eGurleetransf2n})  
with the covariant Dirac (Fock-Ivanenko) equation (\ref{diraccurved2}).

The matrix $S$ is anti-periodic 
$
S(\s,\theta+2\pi) = -S(\s,\theta)
$
and therefore  well-defined only on the double covering of the de Sitter hyperboloid.
The map $(t,\theta)\to S(\s,\theta)$  is thus a map from the double covering of the de Sitter  spacetime $\widetilde{dS_2}$  
with values in the spin group $Sp(1,2)$ (see  Eq. (\ref{Sp})); 
when acting on a spinor field it changes its periodicity: periodic (R) fields become anti-periodic (NS) and viceversa.

The group element  $S(t,\theta) =S(\tilde X(t, \theta))= S(\tilde X)$ 
has a very simple geometrical interpretation that is made clear by examining 
the Lorentz transformation associated to it through the projection (\ref{lorentzmap}):
\begin{equation}
\Lambda(S(t,\theta)) = \left(
\begin{array}{ccc}
 \sec t & -\sin \theta  \tan t & -\cos \theta  \tan t \\
 0 & \cos \theta  & -\sin \theta  \\
 -\tan t & \sec t \sin \theta  & \cos \theta \sec t \\
\end{array}
\right)
\end{equation}
so that 
\beq \Lambda(S(t,\theta)) X(t,\theta) =X(0,0)=\left( \begin{array}{c}
 0 \\
0 \\
1\\
\end{array}
\right)
\eeq
All the above features are not present  in the original construction by G\"ursey   and Lee which was relative to the four-dimensional case.

\section{Cocyclic covariance of the de Sitter Dirac-Fock-Ivanenko field. } \label{cocycle}
Let us apply the map (\ref{mapphi0}) to the field $\Psi_\nu$ defined in Eq. (\ref{tpf2}) and get a quantum 
field $\phi_\nu$ solving the  standard Dirac (Fock-Ivanenko) equation. The  field  $\phi_\nu$ has the NS antiperiodicity 
and therefore is well-defined only on the  manifold $\widetilde{dS_2}$.
Eq. (\ref{trasI}) tells us how the 
field  $\phi_\nu$ is transformed by the action of the de Sitter group: 
%
%
\begin{eqnarray}
 \phi'(\X) 
 &=& 
\Sigma(g,\X) \, \phi(g^{-1} \X), \label{nonlin}
\end{eqnarray}
where the matrix 
\begin{eqnarray}
\Sigma(g,\X) =  S(\X)  \, g\, S({g^{-1}\X})^{-1} 
\end{eqnarray}
is also an element of the spin group $Sp(1,2)$ depending on the point $\X\in \widetilde{dS_2}$ and the group element $g$; 
one immediately verifies that $\Sigma(g,\X)$
is a nontrivial cocyle of $Sp(1,2)$:
\begin{eqnarray}
\Sigma(g_1,\X)  \Sigma(g_2,g_1^{-1}\X)= \Sigma(g_1g_2,\X).
\end{eqnarray}
The de Sitter covariance of the de Sitter Dirac NS field $\phi_\nu$ is thus expressed by Eq. (\ref{nonlin}). 
On the other hand there is no covariant Dirac field (in the above sense) in the Ramond sector.
The following remarkable result play an important technical role in the construction of the de Sitter - Thirring model (presented in a companion paper):

\vskip 5pt 
 \noindent {\em For any $g$ in the spin group $Sp(1,2)$ the cocycle $\Sigma(g,\X)$ is diagonal.}
\vskip 5pt 
We give a proof of the above statement by explicitly exhibiting $\Sigma(g,\X)$ for the one-parameter subgroups of the Iwasawa decomposition (\ref{iwadec}). 
Let us first parametrize  the  matrix $S(\X)$  with the Iwasawa coordinates $(\lambda , \zeta)$ of the point $\X$:
\bea
S(\lambda,\zeta)=(1+\lambda^2)^{\frac14} \left(
\begin{array}{cc}
 {\cos \left(\frac{\zeta }{2}+\arctan{\lambda}\right)} & {-i \sin  \left(\frac{\zeta }{2}+\arctan{\lambda}\right)}\\
 -{i \sin \frac{\zeta  }{2}} & { \cos \frac{\zeta  }{2}} \\
\end{array}
\right).
\eea

\noindent 1) For every spatial rotation $g = k$ and every $\X\in \widetilde {dS_2}$ the cocycle $\Sigma(k,\X)=1$, . 

\noindent 2) For a transformation $a(\kappa)$ belonging to the abelian subgroup $A$ we have
\bea
\Sigma(a(\kappa),\X)= \left(
\begin{array}{cc}
e^{-\chi /2} \frac{{(1+{\lambda '}^2)^\frac 14}\ \sin \left(\frac{\zeta '}{2}\right) }{(1+{\lambda }^2)^\frac 14 \ \sin \left(\frac{\zeta }{2}\right)} & 0 \\
 0 &  e^{\chi /2} \frac{(1+{\lambda }^2)^\frac 14 \ \sin \left(\frac{\zeta }{2}\right)}{{(1+{\lambda '}^2)^\frac 14}\ \sin \left(\frac{\zeta '}{2}\right) }  
\end{array}
\right)
\eea
where
\begin{eqnarray}
\left\{ \begin{array}{l}  \lambda' = \lambda' (-\kappa) = \lambda  \cosh \kappa-\sinh \kappa (\lambda  \cos \a+\sin \a), \\ 
\cot\frac {\a'} 2= \cot\frac {\a'(-\kappa)} 2= e^{-\kappa}\cot \frac \a 2 .
\end{array}\right.\label{trasf2}
 \end{eqnarray}
\noindent 3) For a transformation $n(\mu)$ belonging to the upper triangular subgroup $N$ we have
\begin{equation}
\Sigma(n(\mu),\X)=\left(
\begin{array}{cc}
 \frac{{(1+{\lambda '}^2)^\frac 14}\ \sin \left(\frac{\zeta '}{2}\right) }{(1+{\lambda }^2)^\frac 14 \ \sin \left(\frac{\zeta }{2}\right)} & 0 \\
 0 &   \frac{(1+{\lambda }^2)^\frac 14 \ \sin \left(\frac{\zeta }{2}\right)}{{(1+{\lambda '}^2)^\frac 14}\ \sin \left(\frac{\zeta '}{2}\right) }  
\end{array}
\right)
\end{equation}
where
\begin{eqnarray}
\left\{ \begin{array}{l}  \lambda' = \lambda' (-\mu) = \frac{1}{2} \left(\mu   (2 \lambda -\mu ) \sin \zeta -\mu   (\lambda  \mu +2)\cos \zeta +\lambda  \left(\mu ^2+2\right)\right) \\ 
\cot\frac {\a'} 2= \cot\frac {\a'(-\mu)} 2= \cot \frac \a 2 +\mu.
\end{array}\right.\label{trasf3}
 \end{eqnarray}

\section{Massless fields: from the  the de Sitter manifold to the cylinder and back}
In this concluding section we examine the various incarnations  of the massless Dirac field.
As explained in Appendix \ref{GL}, in the massless case the dressing is simpler (see Eq. (\ref{mapphimassless})): 
\bea && \psi(\s,\theta) =  \frac 1 { \sqrt {\cos t}} S(\s,\theta)  \Psi_0(\s,\theta) \label{transform}
\eea
The LHS has to be understood as a  Dirac field on the double covering of the 
cylinder obtained from the massless field   Eq. (\ref{tpf2masslessbis}).
Computing the two-point function we get
 \begin{eqnarray}
 \frac {S(\s,\theta) (\Omega,  \Psi_0(\s,\theta) \overline\Psi_0(\s',\theta') \Omega) S(\s',\theta')^{-1}
} { \sqrt {\cos t}\sqrt {\cos t'}}=  - \frac { i }{4\pi }\left(
\begin{array}{cc}
 0 &\frac{1}{\sin \left(\frac1{2}({u-u'}) \right)}\\
\frac{1}{\sin \left(\frac1{2}({v-v'}) \right)}& 0 \\
\end{array}
\right)
 \end{eqnarray} 
The two-point function completely characterizes the field: the remarkable result is 
that by the above  construction the covariant massless de Sitter-Dirac field (\ref{tpf20}) is precisely mapped into  the Neveu-Schwarz-Dirac field on the cylinder
\bea && \psi ^{\makebox{\tiny NS}}(\s,\theta)=  \frac 1 { \sqrt {\cos t}} S(\s,\theta)  \Psi_0(\s,\theta)
\eea
and viceversa
\bea &&  \Psi_0(\s,\theta)=   { \sqrt {\cos t}} \, S(\s,\theta)^{-1}  \psi ^{\makebox{\tiny NS}}(\s,\theta). \label{it}
\eea
It is also instructive to apply the inverse transform (\ref{it}) to the two components of the field (\ref{z.100}) separately:
\bea && \Psi_{0,r(l)}(\s,\theta)=   { \sqrt {\cos t}} \, S(\s,\theta)^{-1}  \psi ^{\makebox{\tiny NS}}_{1(2)}(\s,\theta). \label{itlr}
\eea
We get in this way a splitting of the massless Dirac field $\Psi_{0} = \Psi_{0,r}+\Psi_{0,l}$
into its  right and left moving parts (i.e. the parts depending only on the $u$ and $v$ variables):
\begin{eqnarray}
(\Omega,\Psi_{0,r(l)}(\s,\theta) \overline \Psi_{0,r(l)}(\s',\theta') \rangle
=  \frac 1 {4\pi i} (1\pm i \slashed  X) (\slashed  X- \slashed Y)^{-1}
\label{tp228}
\end{eqnarray}
where 
\begin{eqnarray}
\frac 1 {4\pi i} ({\bf 1}+i \slashed  X) (\slashed  X- \slashed Y)^{-1} = A(u,u')= \left(
\begin{array}{cc}
 \frac{\cos \frac{u}{2}\sin \frac{u'}{2}}{4 \pi  \sin \left(\frac{1}{2} (u-u')\right) } &  \frac{i \cos \frac{u}{2}\cos \frac{u'}{2}}{4 \pi  \sin \left(\frac{1}{2} (u-u')\right) } \\
 \frac{i\sin \frac{u}{2}\sin \frac{u'}{2}}{4 \pi  \sin \left(\frac{1}{2} (u-u')\right) } &
 -  \frac{\sin \frac{u}{2}\cos \frac{u'}{2}}{4 \pi  \sin \left(\frac{1}{2} (u-u')\right) }  \\
\end{array}
\right)
\label{tp229}
\end{eqnarray}
\begin{eqnarray}
\frac 1 {4\pi i} ({\bf 1}-i \slashed  X) (\slashed  X- \slashed Y)^{-1} = B(v,v') =  \left(
\begin{array}{cc}
 \frac{\sin \frac{v}{2} \cos \frac{v'}{2}  }{4 \pi  \sin \left(\frac{1}{2} (v-v')\right)} 
 & \frac{i\sin \frac{v}{2} \sin \frac{v'}{2}  }{4 \pi  \sin \left(\frac{1}{2} (v-v')\right)}\\
 \frac{i \cos \frac{v}{2} \cos \frac{v'}{2}  }{4 \pi  \sin \left(\frac{1}{2} (v-v')\right)}&
   -\frac{\cos \frac{v}{2} \sin \frac{v'}{2}  }{4 \pi  \sin \left(\frac{1}{2} (v-v')\right)} \\
\end{array}
\right)
\label{tp229}
\end{eqnarray}
These expression are useful in computing the image of the Ramond field (\ref{z.100a}) under the same transformation:
\bea &&  \Psi^{\makebox{\tiny R}}_0(\s,\theta)=   { \sqrt {\cos t}} \, S(\s,\theta)^{-1}  \psi ^{\makebox{\tiny R}}(\s,\theta). \label{it2}
\eea
The field $\Psi^{\makebox{\tiny R}}_0(\s,\theta)$  is defined on the double covering of the de Sitter spacetime and solves the massless Dirac equation 
(\ref{dSW}) there. Its two-point function is written in the simplest way as follows in terms of the left and right part of the covariant two-point function  as follows:
\begin{eqnarray}
(\Omega , \Psi^{\makebox{\tiny R}}_0(\s,\theta) \overline \psi^{\makebox{\tiny R}}_0(\s',\theta') \Omega)   = 
\cos \left(\frac12 (u-u')\right)A(u,u')+ \cos \left(\frac12 (v-v')\right) B(v,v') \label{rue}
 \end{eqnarray} 
On the other hand, expressing the above two-point function using the ambient space variables gives a very complicated expression, not particularly useful. 
The representation (\ref{rue}) allows to prove
that the $\Psi^{\makebox{\tiny R}}_0(\s,\theta)$  is covariant  under rotations; namely when $g = g(\theta_0)$ is spatial rotation as in (\ref{srot})
 then
\begin{eqnarray}
g(\theta_0) \langle \Psi^{\makebox{\tiny R}}_0(\s,\theta) \overline \psi^{\makebox{\tiny R}}_0(\s',\theta') \rangle  g(-\theta_0) = 
(\Omega, \Psi^{\makebox{\tiny R}}_0(\s,\theta+\theta_0) \overline \psi^{\makebox{\tiny R}}_0(\s',\theta'+\theta_0) \Omega_0) 
 \end{eqnarray} 
On the other hand, the boosts are broken.

\newpage
\appendix
\section{Derivation of the Gursey and Lee's dressing} \label{GL}
The Dirac operator can be rewritten as follows
\beq
iD= {{1}\over{2}}\gamma^{\alpha}\gamma^{\beta}M_{\alpha\beta}=
{\slashed X}{\slashed P}- (X P)
\eeq
where $P_{\alpha}=-i\partial_{\alpha}$ and $X\cdot P= (XP) $. Note that, while the operator $iD$ does not include derivatives w.r.t. the radial variable $r$ the operators at the RHS separately do. 
The important on-shell identity for what follows is the  anticommutator:
\beq \left\{iD,{\slashed X}\right \} = -2i { \slashed X}. \label{ac}
 \eeq
Let  $\Psi$ be a solution of the Dirac equation (\ref{dSW}) and define
\beq \chi =  \frac 1 {\sqrt 2} \left(1-{\slashed X}\right)\Psi.\eeq 
eq. (\ref{ac}) implies that
\bea
\left[iD+i+\nu \right]\left(1+{\slashed X}\right)\chi=
i \left(1-{ \slashed X}\right)\chi + \nu \left(1+{ \slashed X}\right)\chi +\left(1-{ \slashed X}\right) iD\chi =0
 \label{eq2}
\eea
By multiplying the left of the RHS by the operator $(1- {\slashed X } )$ and taking into account that $\chi$ does not depend on the radial variable  we get
\bea
- i { \slashed X} \chi +  \nu \chi - { \slashed X}  iD\chi = 
- i \beta^a \partial_a \chi+ i \beta^2 \chi +  \nu \chi  = 0.
 \label{eq2}
 \eea
The latter identity is known as the Gursey and Lee's equation. 
Let us now define 
\beq
\phi  =  S(t,\theta) \chi .
\label{reff}
\eeq
The spinor $\phi$ satisfies the following equation: 
\bea
- i \alpha^a \partial_a \phi  - i \alpha^a S\left(\partial_a S^{-1}\right)\, \chi  +  i \gamma^2  \phi +  \nu \phi = 0 
\eea
Using the identity 
\bea
 \alpha^t  \alpha_\theta + \alpha^\theta  \alpha_t = -2i\gamma^2
\eea
we may rewrite it as follows:
\bea
  i  \alpha^t \left(\partial_t  - (\partial_t S) S^{-1}- \frac i 2 \alpha_\theta \right)\phi   +    i  \alpha^\theta \left(\partial_\theta     - (\partial_\theta S) S^{-1}- \frac i 2 \alpha_t \right)\phi  - \nu \phi  = 0
 \label{eGurleetransf2} 
\eea
Finally, the identifications
\bea 
- (\partial_t S) S^{-1}- \frac i 2 \alpha_\theta =0= \Gamma_\s , \ \ \        - (\partial_\theta S) S^{-1}- \frac i 2 \alpha_t =\left(
\begin{array}{cc}
 -\frac1 {2}{\tan \s} & 0 \\
 0 & \frac 1 {2} {\tan \s} \\
\end{array}
\right)=\Gamma_\theta,\cr
\eea
show that Eq. (\ref{eGurleetransf2}) coincides  with the covariant Dirac (Fock-Ivanenko) equation (\ref{diraccurved2}). 
The above construction more or less coincide with  the original proposal by Gursey and Lee.
We may also proceed the other way around and learn something. Let  again $\Psi$ be a solution of the Dirac equation (\ref{dSW}) and define first
\beq \zeta=  S(t,\theta) \Psi.
\eeq 
Then
\bea
S \left[iD+i+\nu \right] S^{-1} \zeta=
-\gamma^2 S \slashed P S^{-1}  \zeta +(i+\nu) \zeta=0
\eea
or else 
\bea
 S \slashed P S^{-1}  \zeta  +(i+\nu) \gamma^2\zeta= 
-  i  \alpha^i \left(\partial_i  +\Gamma_i \right)\zeta   +    \nu \gamma^2 \zeta  = 0
 \label{eGurleetransf2} 
\eea
where again we took into account the fact that $\zeta$ does not depend on $r$.
By defining
\bea  \zeta= \frac 1 {\sqrt 2}(1-\gamma^2)  \phi \eea 
we get again
\bea
 (1+\gamma^2)\left[i \alpha^i \left(\partial_i  +\Gamma_i \right)\phi  -    \nu \phi \right] = 0
 \label{eGurleetransf2} 
\eea
This second step is unnecessary in the massless case. Summarising, given a solution of the Dirac  equation (\ref{dSW}) the spinor 
\bea
\phi (t,\theta)= \frac 1 {\sqrt 2}  S(t,\theta) (1- \slashed  X )\Psi (t,\theta) = \frac 1 {\sqrt 2}  (1+  \gamma^2 ) S(t,\theta) \Psi(t,\theta)
 \label{mapphi} 
\eea
solves the covariant Dirac (Fock-Ivanenko) equation (\ref{diraccurved2}) (with $m=\nu$).  The inverse map is
\bea
\Psi (t,\theta)= \frac 1 {\sqrt 2}   (1+ \slashed  X ) S(t,\theta)^{-1}\phi (t,\theta) = \frac 1 {\sqrt 2}  S(t,\theta)^{-1} (1-  \gamma^2 ) \phi(t,\theta)
 \label{mapphi2} 
\eea
In the massless case the above maps reduce to
\bea
\phi_0 (t,\theta)= S(t,\theta) \Psi_0 (t,\theta)   \label{mapphimassless} 
\eea

Taking into account the conjugacy relation $\gamma^0 S^+\gamma^0=S^{-1}$ (defining our realisation of $Sp(1,2)$) we also have
\beq \overline \phi(\s,\theta) = \phi^+(\s,\theta) \gamma^0 
= \frac 1 {\sqrt 2} \overline {\Psi}(\s,\theta) (1+\slashed  X) S(\s,\theta)^{-1} = \frac 1 {\sqrt 2} \overline {\psi}(\s,\theta) S(\s,\theta)^{-1}(1-\gamma^2). \eeq 

\section{Maximally analytic vacua} \label{BB}
We summarise here a small part of the scalar harmonic analysis on the de Sitter space
(see e.g. \cite{bm}) used in the  text. The main result is the following:
for ${z} \in \TT_-$ and ${z'} \in \TT_+$ in $d$ spacetime dimensions one has
\begin{eqnarray}
&& \int_{S_0}(\xi\cdot {z})^{1-d-\lambda}(\xi\cdot {z'})^{\lambda}\,
d\xx = \int_\gamma(\xi\cdot {z})^{1-d-\lambda}(\xi\cdot {z'})^{\lambda}\,
\formdif(\xi)=
\label{r.2}\\
&&={2\pi^{d\over 2}e^{i\pi\left ( \lambda+{d-1\over 2} \right )}\over
\Gamma \left ({d\over 2}\right )}
F\left(-\lambda,\ \lambda+d-1;\ {d\over 2};\ {1-\zeta\over 2}\right ),
\ \ \ \zeta = {z}\cdot {z'}, \ \ \ \lambda\in {\bf C}\, .
\label{r.2.1}\end{eqnarray}
Here $\gamma$ is any smooth $(d-1)$-cycle homotopic to the unit sphere 
$S_0$ in $C_+\setminus \{0\}$ and $\formdif$ is the $(d-1)$-differential form obtained from the rotation invariant measure $d\xi$ on the sphere.
Both sides of (\ref{r.2}) are entire in $\lambda \in \bC$,
as well as the rhs of (\ref{r.2.1}).
If $\lambda$ is not a pole of $\Gamma(-\lambda)\Gamma(\lambda+d-1)$,
${z}\in\TT_-$, ${z'}\in\TT_+$, we denote
\begin{eqnarray}
 \wl({z}\cdot {z'})  &=& c(\lambda)
\int_{S_0}(\xi\cdot {z})^{1-d-\lambda}(\xi\cdot {z'})^{\lambda}\,
d\xx,
\label{r.3}\\
c(\lambda) &=& {\Gamma(-\lambda)\Gamma(\lambda+d-1)
e^{-i\pi\left ( \lambda+{d-1\over 2} \right )}\over 2^{d+1}\pi^d}\ .
\label{r.4}
\end{eqnarray}
Two equivalent
expressions for $\wl$:
\begin{eqnarray}
\wl({z}\cdot {z'}) &=& {\Gamma(-\lambda)\Gamma(\lambda+d-1)\over
(4\pi)^{d/2}\Gamma\left({d\over 2}\right)}
F\left(-\lambda,\ \lambda+d-1\ ;\ {d\over 2};\ {1-\zeta\over 2}\right )
\label{r.0}\\
&=&
{\Gamma(-\lambda)\Gamma(\lambda+d-1)\over 2 (2\pi)^{d/2}}
(\zeta^2-1)^{-{d-2 \over 4}}\,P_{\lambda +{d-2\over 2}}^{-{d-2 \over 2}}(\zeta),
\ \ \ \zeta = {z}\cdot {z'}\ .
\label{r.1}\end{eqnarray}

\section*{Acknowledgments} U.M thanks the Institut des Hautes Etudes Scientifiques (Bures-sur-Yvette) for its hospitality and support.

  \end{document}

  \section{Transformation 4 dim}
If $\psi$ is a solution of the Dirac equation (\ref{dSW}) the spinor
\beq
\chi= \frac{1- \beta^{4}}{\sqrt 2}\psi
\eeq
satisfies the G\"ursey--Lee equation and the spinor 
\beq
\phi = S \chi = \frac 1 {\sqrt 2}  S (1- \beta^{4}){\psi} = \frac 1 {\sqrt 2}  S (1+\slashed x){\psi}
\eeq
satisfies the standard covariant Dirac equation on the de Sitter manifold. Since  $\gamma^0 S^+\gamma^0=S^{-1}$ the conjugate spinor is given by 
\beq \overline \phi = \phi^+ \gamma^0 = \frac 1 {\sqrt 2} \overline {\psi} (1+\slashed x) \gamma^0 S^+\gamma^0 = \frac 1 {\sqrt 2} \overline {\psi} (1+\slashed x) S^{-1}.
\eeq
The two point function of the transformed field is then easily computed. \\
Since $S_x \slashed x S_x^{-1} =-\gamma^4$ we get
\begin{eqnarray}
\langle \phi (x)\overline \phi(y)\rangle  &= &\frac 1 {2} S_x (1+\slashed x )  \langle \psi (x)\overline {\psi}(y)\rangle (1+\slashed y )  S_y^{-1}
\cr&= & \frac{i}{4\pi^2}\ S_x (1+\slashed x )  
\frac{({\slashed x}-{\slashed y)}}{((x-y)^{2})^2} (1+\slashed y )  S_y^{-1} \cr &=&  \frac{i}{2\pi^2}\ S_x 
\frac{({\slashed x}-{\slashed y)}}{((x-y)^{2})^2}  S_y^{-1}  = \frac{1}{2\pi^2 i {((x-y)^{2})^2}} \left[ \gamma^4, \, S_x S_y^{-1} \right]
\end{eqnarray}
Finally
\begin{eqnarray}
\langle \phi (x)\overline \phi(y)\rangle =  \frac{i}{4\pi} \sqrt{\cos (s) \cos (s')} \left(
\begin{array}{cc}
 0 &\frac 1{\sin\left(\frac{1}{2} (s-s'+\theta -\theta' )\right) }\\
  \frac 1 {\sin \left(\frac{1}{2} (s-s'-\theta +\theta' )\right)} & 0 \\
\end{array}
\right)\cr 
= 
\frac{1}{2\pi} \sqrt{\cos (s) \cos (s')} \sum_{n=0}^\infty  \left(
\begin{array}{cc}
 0 &   e^{\frac{1}{2} i (2 n+1) (s-s'+\theta -\theta' )} \\
e^{\frac{1}{2} i (2 n+1) (s-s'+\theta -\theta' )} & 0 \\
\end{array}
\right)\cr    
 \end{eqnarray} \newpage
 \section{4D}
 $$ \left(
\begin{array}{cccc}
 \frac{e^{-\frac{1}{2} i (\theta +\psi )} \cos \left(\frac{s}{2}\right) \cos \left(\frac{\phi }{2}\right)}{\sqrt{\cos (s)}} & -\frac{i e^{-\frac{1}{2} i
   (\theta -\psi )} \cos \left(\frac{s}{2}\right) \sin \left(\frac{\phi }{2}\right)}{\sqrt{\cos (s)}} & \frac{i e^{\frac{1}{2} i (\theta -\psi )} \cos
   \left(\frac{\phi }{2}\right) \sin \left(\frac{s}{2}\right)}{\sqrt{\cos (s)}} & \frac{e^{\frac{1}{2} i (\theta +\psi )} \sin \left(\frac{s}{2}\right)
   \sin \left(\frac{\phi }{2}\right)}{\sqrt{\cos (s)}} \\
 -\frac{i e^{\frac{1}{2} i (\theta -\psi )} \cos \left(\frac{s}{2}\right) \sin \left(\frac{\phi }{2}\right)}{\sqrt{\cos (s)}} & \frac{e^{\frac{1}{2} i
   (\theta +\psi )} \cos \left(\frac{s}{2}\right) \cos \left(\frac{\phi }{2}\right)}{\sqrt{\cos (s)}} & \frac{e^{-\frac{1}{2} i (\theta +\psi )} \sin
   \left(\frac{s}{2}\right) \sin \left(\frac{\phi }{2}\right)}{\sqrt{\cos (s)}} & \frac{i e^{-\frac{1}{2} i (\theta -\psi )} \cos \left(\frac{\phi
   }{2}\right) \sin \left(\frac{s}{2}\right)}{\sqrt{\cos (s)}} \\
 -\frac{i e^{-\frac{1}{2} i (\theta +\psi )} \cos \left(\frac{\phi }{2}\right) \sin \left(\frac{s}{2}\right)}{\sqrt{\cos (s)}} & -\frac{e^{-\frac{1}{2} i
   (\theta -\psi )} \sin \left(\frac{s}{2}\right) \sin \left(\frac{\phi }{2}\right)}{\sqrt{\cos (s)}} & \frac{e^{\frac{1}{2} i (\theta -\psi )} \cos
   \left(\frac{s}{2}\right) \cos \left(\frac{\phi }{2}\right)}{\sqrt{\cos (s)}} & -\frac{i e^{\frac{1}{2} i (\theta +\psi )} \cos \left(\frac{s}{2}\right)
   \sin \left(\frac{\phi }{2}\right)}{\sqrt{\cos (s)}} \\
 -\frac{e^{\frac{1}{2} i (\theta -\psi )} \sin \left(\frac{s}{2}\right) \sin \left(\frac{\phi }{2}\right)}{\sqrt{\cos (s)}} & -\frac{i e^{\frac{1}{2} i
   (\theta +\psi )} \cos \left(\frac{\phi }{2}\right) \sin \left(\frac{s}{2}\right)}{\sqrt{\cos (s)}} & -\frac{i e^{-\frac{1}{2} i (\theta +\psi )} \cos
   \left(\frac{s}{2}\right) \sin \left(\frac{\phi }{2}\right)}{\sqrt{\cos (s)}} & \frac{e^{-\frac{1}{2} i (\theta -\psi )} \cos \left(\frac{s}{2}\right)
   \cos \left(\frac{\phi }{2}\right)}{\sqrt{\cos (s)}} \\
\end{array}
\right)$$